\documentclass[a4paper,letterpaper]{IEEEtran}
\usepackage{amsmath,amssymb,epsfig,enumerate,cite,footnote,authblk,color,hyperref}
\hypersetup{
    colorlinks,%
    citecolor=blue,%
    filecolor=blue,%
    linkcolor=blue,%
    urlcolor=blue
}
\definecolor{bluecolor}{rgb}{0,0.,1.}

\definecolor{redcolor}{rgb}{.7,0.,0.}

\newcommand{\pr}[1]{\left( #1\right)}

\newcommand{\es}[1]{\begin{equation}\begin{split}#1\end{split}\end{equation}}
\newcommand{\est}[1]{\begin{equation*}\begin{split}#1\end{split}\end{equation*}}

\newcommand{\rr}{\mathbf{r}}
\newcommand{\dd}{\textrm{d}}

\providecommand{\U}[1]{\protect\rule{.1in}{.1in}}
\pdfoutput=1
\providecommand{\U}[1]{\protect\rule{.1in}{.1in}}

\newtheorem{lemma}{Lemma}
\newcommand{\clred}[1]{{\color{black} #1}}

\begin{document}

\title{Connectivity Scaling Laws in Wireless Networks}
\author{Justin P. Coon, Orestis Georgiou, and~Carl P.~Dettmann
  \thanks{J.~P.~Coon is with the Department of Engineering Science, University of Oxford, Parks Road, Oxford, UK, OX1 3PJ.} 
  \thanks{O. Georgiou is with the Toshiba Telecommunications Research Laboratory, 32 Queen Square, Bristol, UK, BS1 4ND.}
  \thanks{C. P. Dettmann is with the School of Mathematics, University of Bristol, University Walk, Bristol, UK, BS8 1TW.}
}

\maketitle

\begin{abstract}
We present scaling laws that dictate both local and global connectivity properties of bounded wireless networks.  These laws are defined with respect to the key system parameters of per-node transmit power and the number of antennas exploited for diversity coding and/or beamforming at each node.  We demonstrate that the local probability of connectivity scales like $\mathcal{O}(z^\mathcal C)$ in these parameters, where $\mathcal C$ is the ratio of the dimension of the network domain to the path loss exponent, thus enabling efficient boundary effect mitigation and network topology control.
\end{abstract}

\begin{keywords}
Connectivity, Outage, boundaries, MIMO.
\end{keywords}

\section{Introduction}

Ad hoc wireless networks have found use in a plethora of applications ranging from environmental monitoring to vehicle-to-vehicle communications primarily due to their ability to increase coverage through multihop transmissions, and to autonomously organise and initiate communications in a decentralised manner.
These networks possess commonality insomuch as the number and spatial distribution of nodes in the network are often random which is also why they are usually modelled as random geometric graphs \cite{penrose2003random}, granting access to theoretical analysis of network performance and ultimately engineering insight.
In practice, this understanding can lead to improved protocols and network deployment strategies \cite{younis2008strategies}, for example, understanding how transmission range affects the underlying network topology can reduce the cost of distributed algorithms, save energy, and lower interference between nodes while maintaining high levels of network connectivity \cite{santi2005topology} usually measured as the probability that an ad hoc network is \emph{fully }connected, i.e. there exists at least one reliable multihop path connecting every two nodes in the network. 

For dense networks, several analytical results on connectivity have been published particularly in the form of insightful scaling laws. For example, in \cite{Gupta1998}, the authors derive a power scaling law that ensures full connectivity is achieved \emph{almost surely} as the number of nodes in the network tends to infinity. 
In \cite{Xue2004}, the number of nearest neighbours required to achieve full connectivity asymptotically is studied.

While early works considered unbounded (infinite) networks, more recent research has attempted to better quantify the role that boundaries play in finite networks. 
Simple confining geometries (e.g., cubes, spheres, rectangles) were studied in~\cite{Jia2006,Mao2012,Khalid2014,estrada2015random}.  
A more versatile framework based on a cluster expansion approach was recently detailed by the authors in~\cite{Coon2012a} capable of treating more complicated geometrical network domains (convex polytopes) and encompasses several aspects of subsequently reported theories (cf.~\cite{Khalid2014}). The framework has also been shown to yield more accurate results than conventionally accepted approximations in some cases~\cite{Khalid2014} and may even accommodate directional antenna gains \cite{georgiou2013connectivity}.

In this letter we build on the framework presented in \cite{Coon2012a,georgiou2013connectivity} and derive for the first time connectivity scaling laws (see equations \eqref{eq:MPT}, \eqref{Mdc} and \eqref{eq:Mb}) with regards to transmit power, and the number of antennas employed by transceiver pairs as well as the adopted transmission scheme: a) orthogonal space-time block coding (STBC) and b) beamforming (MIMO with maximum ratio combining - MIMO-MRC).
We show that the local network connectivity scales like $\mathcal{O}(z^\mathcal C)$ in terms of system parameters $z$, where $\mathcal C$ is the ratio of the dimension of the network domain $d$ to the path loss exponent $\eta$.
We conclude by discussing the merits of these transmission schemes with respect to antenna diversity gain, how the derived scaling laws can be used to mitigate boundary effects and how our work can be adapted to account for interference limited networks \cite{location,Haenggi2009}.



\section{Network Definitions and System Model}
\label{sec:net_conn}

\textit{1) Node Deployment:}
Consider a network of $N$ uniformly distributed nodes in a $d$-dimensional convex domain $\mathcal{V}\subseteq\mathbb{R}^{d}$ with location coordinates
$\mathbf{r}_{i}\in\mathcal{V}$ for $i=1,2,\ldots,N$. 
This is equivalent to a Binomial Point Process (BPP) with node density $\rho= N/V$, where $V= |\mathcal{V}|$, and is a common configuration for modelling ad hoc networks in confined geometries. 

\textit{2) Path-loss and Fading:}
The signal-to-noise ratio (SNR) is a commonly used metric to quantify the quality and reliability of a communication link. 
Due to path loss the signal power received by a destination node is inversely proportional to the separation distance $r_{ij}=|\rr_i - \rr_j|$ between two nodes located at $\rr_i$ and $\rr_j$, and can be modelled by 
\es{
g(r_{ij})= (\epsilon +r_{ij}^\eta)^{-1}, \qquad \eta\geq 2, \quad \epsilon > 0
}
where $\eta$ is the path loss exponent and $\epsilon \ll 1$ ensures that the attenuation function $g(r_{ij})$ is non-singular at zero distance.

Due to small-scale fading, it is a standard assumption that the coefficient $h_{kl}\in \mathbb{C}$ modelling the transfer characteristics of the MIMO channel between the $k$th transmit antenna $(1\leq k \leq m)$ and the $l$th receive antenna ($(1\leq l \leq n)$) of nodes $i$ and $j$ respectively is a circularly symmetric complex Gaussian random variable with zero mean and unit variance. 
Consequently, for $m=n=1$ the channel gain $X_{ij}$ between nodes $i$ and $j$ is an exponentially distributed random variable $X_{ij} =|h_{1,1}|^2\sim \exp(1)$ corresponding to the usual SISO Rayleigh fading model adopted in most wireless ad hoc network literature due to its mathematical tractability.

\textit{3) Pairwise Connectivity:}
Assuming  negligible  inter-node  interference (e.g.  perfect CDMA/TDMA) and lossless 
antennas, we define the pairwise connectivity through the relation 
\es{
H_{ij} = P(\textrm{SNR}_{ij} \geq \wp)
\label{H1}}
i.e. the complement of the outage probability, where  the average received signal-to-noise ratio is given by $\textrm{SNR}_{ij}=  g(r_{ij})X_{ij} / \beta$, and the parameter $\beta \propto P_T^{-1}$ depends on transmit power $P_T$, the center frequency of the transmission and the power of the long-time average background noise at the receiver ($\beta$ defines the length scale).
We therefore have that $H_{ij}(r_{ij}) = 1- F_{X_{ij}}(\wp \beta (\epsilon + r_{ij}^\eta))$,
where $F_{X_{ij}}$ is the CDF of the channel gain $X_{ij}$.
Consequently, for $m=n=1$ we obtain
$
H_{ij}(r_{ij}) = \exp(-\wp \beta (\epsilon + r_{ij}^\eta)) 
$.

\section{Connectivity Mass and Scaling Laws}

Scaling laws describe how network connectivity properties (both local and global) scale with various network parameters.

\textit{1) Local Connectivity:}
A good measure of the local connectivity of a transmitter node located at $\rr_i$ is given by the spatial average of $H_{ij}$ over all possible receiver positions 
\es{
M_i(\rr_i) = \int_\mathcal{V} H_{ij}(r_{ij}) \dd \rr_j 
\label{MM}}
defining the likelihood that a node located at $\rr_i$ will connect (i.e. has SNR greater than $\wp$) to some other arbitrary node in the network domain $\mathcal{V}$.
Note that $(N-1) M_i(\rr_i)/V$ is the expected degree of node $i$.
In fact $M_i$ is related to a number of local and global network observables, and for that reason we call $M_i$ the \textit{connectivity mass} \cite{georgiou2013connectivity}.

\textit{2) Global Connectivity:}
A global measure of network connectivity is the probability that every node can communicate with every other node in a multihop fashion. 
This is captured by the notion of full connectivity $P_{fc}$ which at high node densities $\rho$ can be expressed as (see \cite{Coon2012a,georgiou2013connectivity} for more details)
\es{
P_{fc} \approx 1- \rho \int_\mathcal{V} e^{- \rho M_i(\rr_i) } \dd \rr_i
\label{Pfc}}
essentially saying that full connectivity is the complement of the probability of an isolated node.
The connection between $P_{fc}$ and $M_i$ is clear from \eqref{Pfc}: as $M_i$ increases, the probability of an isolated node decreases and hence $P_{fc}$ monotonically increases to $1$.
Moreover, in the dense regime of $\rho\to \infty$ we have that the integral in \eqref{Pfc} will be dominated by integration regions where $M_i$ is small, i.e. near the boundaries of $\mathcal{V}$.

\textit{3) Boundary effects:}
Taylor expanding $M_i$ at $\mathbf r_i$ situated on the boundary of $\mathcal{V}$ yields the leading-order expression\footnote{Assuming that $\mathbb E[X_{ij}^{\mathcal C}] <\infty$, with $\mathbb E[\cdot]$ denoting the expectation operator, the integral in~\eqref{M} is bounded.
Moreover, the upper limit of integration in \eqref{M} is justified if $H_{ij}$ decays exponentially with increasing distance $x_{ij}$ and $\mathcal{V}$ is larger than the typical transmission range. 
}
\es{
M_i &\approx \omega \int_0^\infty \!\! r^{d-1} \pr{1- F_{X_{ij}}(\wp \beta (\epsilon + r^\eta))} \dd r \\
&=   \frac{\omega}{\eta (\wp \beta)^\mathcal{C}} \int_{\wp \beta \epsilon}^\infty \!\! (x-\wp \beta \epsilon)^{\mathcal{C}-1} \pr{1- F_{X_{ij}}(x)} \dd x
\label{M}
}
where we define $\mathcal C = d/\eta$ to be the \emph{connectivity exponent} and $\omega = \int \mathrm d \Omega$ is the solid angle as seen from $\mathbf r_i$, with $\Omega = 2\pi^{d/2}/\Gamma(d/2)$ being the full solid angle in $d$ dimensions and $\Gamma(\cdot)$ is the standard gamma function. 
Equation \eqref{M} implies that the connectivity mass $M_i$ of node $i$ is proportional to the available angular area $\omega$ for other nodes to connect to.
For example if $\rr_i$ is located at the corner of a 2D regular $n$-gon, then $\omega=\pi(1-2/n)$.
Hence, boundary effects have a direct (negative) impact on both local and global network properties through $\omega$ motivating the study of how one can mitigate such effects by means of network design parameters.

\textit{4) Power law scaling:}
Taking a leading order approximation of \eqref{M} for small $\epsilon$ we have that
\es{
M_i &\approx \frac{\omega}{ \eta (\wp \beta)^\mathcal{C}}  \int_0^\infty \!\! x^{\mathcal{C}-1} \pr{1- F_{X_{ij}}(x)} \dd x  + \ldots
\label{M1}
}
which suggests that we can write the following scaling law
\begin{equation}\label{eq:MPT}
  M_i \approx K_1 P_T^{\,\mathcal C}
\end{equation}
with $K_1$ independent of the transmit power $P_T$.
Correction terms to \eqref{M1} can be shown to be $\mathcal{O}(\epsilon^{\min(\mathcal{C},1)})$ but are omitted for the sake of brevity. 
Note that for $m=n=1$ it follows that $M_i \approx \frac{\omega}{ \eta (\wp \beta)^\mathcal{C}}\Gamma(\mathcal{C})$.
Significantly, since $M_i$ is proportional to the mean degree of node $i$ it follows that doubling the transmit power will double the expected number of receivers whose SNR level is higher than $\wp$, if $\mathcal{C}=1$.
It is particularly interesting to note the conditions under which power scaling provides a progressive improvement to local connectivity, versus those conditions under which diminishing returns are experienced with an increase in $P_T$. 
For example, a high-dimensional network ($d = 3$) operating in low path loss conditions $\eta<3$ will benefit from the former behaviour as $P_T$ is increased; however, a low-dimensional network ($d=2$) located in a cluttered environment where high path loss conditions prevail will experience the latter. 

The power law given by \eqref{eq:MPT} provides useful insights into the behaviour of random geometric networks, which can be used to enhance network designs in practice. 
For example, in the case of transmit power scaling \eqref{eq:MPT}, we can use this analysis to deduce that, for some nominal transmit power $P_{T_{0}}$ that defines the target connectivity probability of a homogeneous network, we must choose $P_T$ for the bounded network to satisfy $P_T = (\Omega/\omega)^{1/\mathcal C}P_{T_0}$.
We conclude this discussion by pointing out that the power law~\eqref{eq:MPT} arises from the fact that $M_i \propto \beta^{-\mathcal C}$.  Thus, one can infer that scaling laws in other key system parameters are affected by the connectivity exponent in the same way i.e. $\mathcal{O}(z^\mathcal{C})$ (e.g. frequency or antenna gain).

\section{\clred{Multi-Antenna Systems}}
\label{sec:diversity}

Consider the case where each node in the network possesses $m$ transmit antennas and $n$ receive antennas, and one of two signalling mechanisms is employed: diversity coding following the conventional STBC scheme derived from generalized complex orthogonal designs (GCODs)~\cite{Tarokh1999}, and transmitter/receiver beamforming, also known as MIMO-MRC~\cite{Kang2003}. 

\textit{1) Diversity Coding:}
It is well known that the performance of a point-to-point STBC system is governed by the Frobenius norm of the associated $n \times m$ channel matrix $\mathbf H$ with $h_{kl}$ as its entries, that is the channel gain $X_{ij} = |\mathbf H\|_F^2 = \sum_{k,l}{|h_{kl}|^2}$. 
Consequently, $X_{ij}$ is chi-squared distributed with $2mn$ degrees of freedom, and its cumulative distribution function is given by $F_{X_{ij}}(x) = \gamma(mn,x)/\Gamma(mn)$, where $\gamma(\cdot,\cdot)$ is the lower incomplete gamma function.
Moreover, for $m \geq 1$, the post-processing received SNR can be expressed as $\text{SNR}_{ij}= \frac{\zeta_m}{m} g(r_{ij}) X_{ij}/\beta$, where $\zeta_m = 1$ if $m=1,2$ and $\zeta_m = 2$ otherwise.  The factor of $\zeta_m/m$ arises from power normalization and the fact that the rate of a code derived from GCOD is $\frac{1}{2}$ for $m > 2$~\cite{Tarokh1999}. 
The pairwise connectivity is thus
\es{
H_{ij}= P\pr{ X_{ij} \geq \frac{m\wp\beta}{\zeta_m g(r_{ij})}}= 1-\frac{\gamma\pr{m n , \frac{m \wp \beta }{\zeta_m g(r_{ij})} }}{\Gamma(mn)}
\label{Hdc}}
from which we evaluate the connectivity mass~\eqref{M1} for $\epsilon\ll 1$
\es{
  M_{\text{dc}} \approx \frac \omega d \left(\frac{\zeta_m}{m\wp\beta}\right)^{\mathcal C}\frac{\Gamma(mn+\mathcal C)}{\Gamma(mn)}
\label{Mdc}}
where we have used a ``dc'' subscript to indicate the relation to diversity coding.  For large $m$ and/or $n$, we can use the Stirling formula $\Gamma(x)\sim \sqrt{2\pi}\,x^{x+\frac 1 2} e^{-x}$ to obtain the scaling~law
\begin{equation}\label{eq:Mdc}
  M_{\text{dc}} \approx \frac{\omega}{d}\left(\frac{\zeta_m n}{\wp\beta}\right)^{\mathcal C}\Big(1 + O\Big(\tfrac{1}{mn}\Big)\Big),\quad m,n\rightarrow\infty.
\end{equation}
Note that the connectivity exponent $\mathcal C$ arises naturally in a manner similar to the power scaling law in \eqref{eq:MPT}. 

\textit{2) Beamforming:}
For MIMO-MRC transmissions, the received SNR (after MRC) is $\text{SNR}_{ij}= \lambda_{\max}(\mathbf H^\dagger \mathbf H) g(r_{ij})/\beta$, (i.e. $X_{ij}=\lambda_{\max}(\mathbf H^\dagger \mathbf H)$) with $\lambda_{\max}(\cdot)$ denoting the maximum eigenvalue of the argument~\cite{Kang2003}. 
A closed form expression for $H_{ij}$ like in \eqref{Hdc} is not possible here.
To make progress we look at the behaviour of $\lambda_{\max}$ in the limit of large $m,n$ and apply the following result due to Edelman~\cite[Lemma 4.3]{Edelman1988}\footnote{Edelman's result assumed the complex Gaussian entries of $\mathbf H$ had unit variance \emph{per dimension}, whereas the result we use assumes each entry has unit variance \emph{in total}, and thus this lemma is slightly different from ~\cite{Edelman1988}.}:
\begin{lemma}\label{lem:edelman}
  Let $x_n \xrightarrow{p} x$ signify that for all $\delta > 0$, $\lim_{n\rightarrow\infty}{P(|x-x_n|> \delta)}=0$.  Now, suppose the $n \times m$ matrix $\mathbf H$ has independent circularly symmetric complex Gaussian entries, each with zero mean and unit variance. Then $\mathbf W = \mathbf {H^\dagger H}$ has a complex Wishart distribution and
\begin{equation}
  (1/n)\lambda_{\max}(\mathbf W) \xrightarrow{p} (1+\sqrt y)^2,\qquad \frac m n \rightarrow y,~0\leq y < \infty.
\end{equation}
\end{lemma}
We therefore draw inspiration from this lemma and write 
\begin{equation}
  H_{ij}(r_{ij}) \approx \left\{
    \begin{array}{ll}
  	1, & r_{ij} < \Big(\frac{(1+\sqrt y)^2n}{\wp\beta}\Big)^{\frac{1}{\eta}} \\ 
  	0, & \text{otherwise} \\
    \end{array}
   \right.
\end{equation}
for large $n$.  
Letting $\mu(n) = (1+\sqrt y)^2n$, we write
\begin{equation}
  \frac{M_{\text{b}}}{ \omega} \approx \int_{0}^{ (  \frac{\mu}{\wp\beta} )  ^{\frac{1}{\eta}}}\!\!r^{d-1}   \,\mathrm{d}r + \varepsilon(n) = \frac 1 d \left(\frac{(1+\sqrt y)^2n}{\wp\beta}\right)^{\mathcal C}+ \varepsilon(n)
\end{equation}
where a ``b'' subscript is used to indicate the relation to beamforming.  The error term is given~by
\est{
\varepsilon (  n )  = \int_{ (\frac{\mu}{\wp\beta} )  ^{\frac{1}{\eta}}}^{\infty}r^{d-1}
  H (r )  \,\mathrm{d}r
  - \int_{0}^{ (  \frac{\mu}{\wp\beta} )  ^{\frac{1}{\eta}}}r^{d-1} 
  (1-  H (  r )) \,\mathrm{d}r
}
which grows like $\mathcal{O}(n^{\mathcal C - \frac 2 3})$. We omit the proof of this due to space restrictions. Thus, we arrive at the following scaling law
\es{\label{eq:Mb}
  M_{\text{b}} \approx \frac{\omega}{d} \left(\frac{(1+\sqrt{y})^2 n}{\wp\beta}\right)^{\mathcal C}\left(1 + \mathcal{O}\!\left(n^{-\frac 2 3}\right)\right)
}
valid for the limit of $m,n\rightarrow\infty$ where $m/n \rightarrow y$.

\section{Comparison of the Two Multi-Antenna Schemes}

\begin{figure}[t]
\begin{center}
\includegraphics[width=8.85cm]{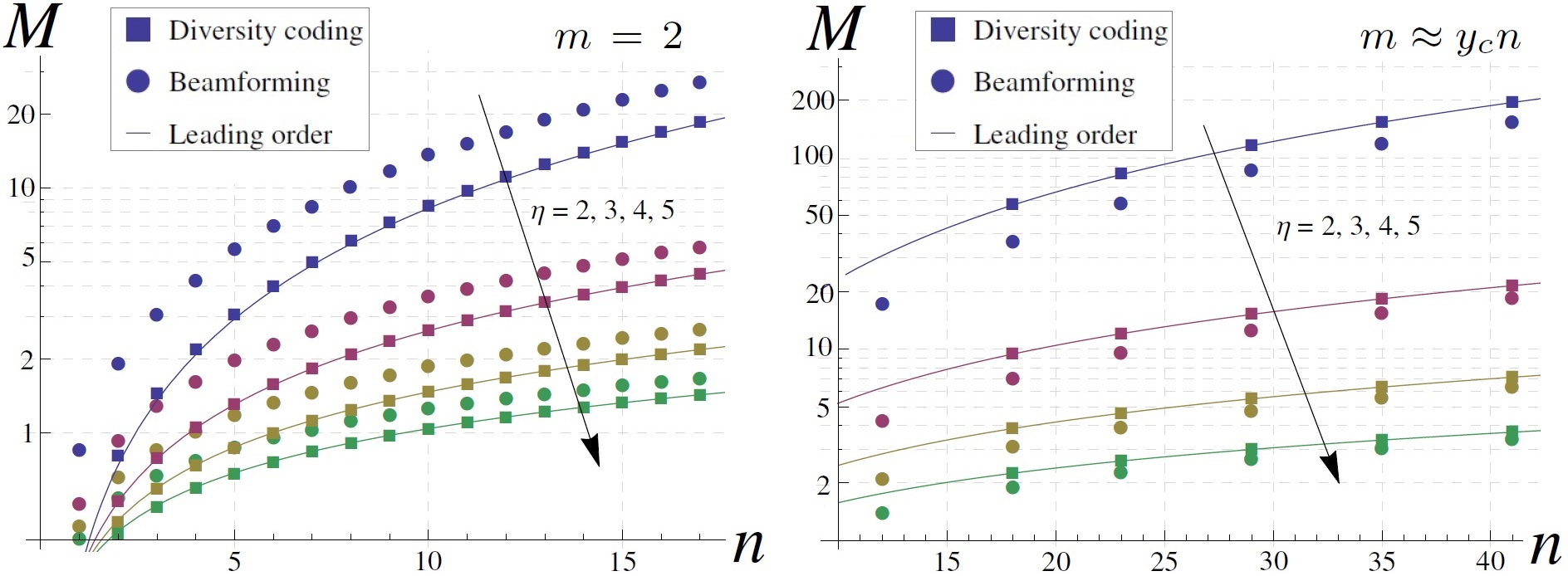}
\caption{Connectivity mass vs.~$n$ for $d = 3$ and $m = 2$ (left panel) and $m\approx y_c n$ (right panel), and various values of $\eta$, corresponding to connectivity exponents $\mathcal C= \frac{d}{\eta} = \frac 3 2, 1, \frac 3 4, \frac 3 5$. The solid lines correspond to the leading-order term given in~\eqref{eq:Mb} (equivalently~\eqref{eq:Mdc}), whereas the data represented by markers was obtained from~\eqref{Mdc} for the diversity coding scenario and by numerically calculating~\eqref{M1} for the beamforming case. 
}
\label{fig1}
\end{center}
\end{figure}

\textit{1) The case of $m=2$:}
Suppose the number of transmit antennas per node is fixed at $m=2$, in which case $\zeta_m = 1$ and $y = 0$, and thus the leading order of $M_{\text{dc}}$ is the same as that of $M_{\text{b}}$ (c.f. \eqref{eq:Mdc} and \eqref{eq:Mb}). However, we see that the first-order corrections for the two observables differ. This is illustrated in the left panel of Fig.~\ref{fig1}, where exact results for the connectivity masses of the two systems are presented along with leading-order terms as a function of $n$.  In the figure, the solid lines show the leading-order behaviour, while the marker data was obtained from the direct calculation of~\eqref{Mdc} for diversity coding and numerically calculating~\eqref{M1} for beamforming. 

Three observations can be made from this example.  The first is that the difference in first-order corrections is apparent, and beamforming actually provides a connectivity benefit over diversity coding for finite numbers of receive antennas i.e. $M_{\text{dc}}  \lesssim M_{\text{b}}$. Yet, convergence to the leading order can be seen for both schemes as $n\to\infty$. 
The second observation is that the leading-order expression well approximates the exact connectivity mass, even for small numbers of antennas, particularly for $M_{dc}$. 
The third observation is that the derivative of the connectivity mass satisfies $M'(n) \propto n^{\mathcal C -1}$ implying that progressive improvements are obtained for $\mathcal C > 1$ and diminishing returns for $\mathcal C < 1$.

\textit{2) The case of $m>2$:}
For any other fixed $m$ greater than two, $M_{\text{dc}} > M_{\text{b}}$ to leading order by a factor of $2^{\mathcal C}$.  However, STBC suffers from a lower rate than MIMO-MRC in this case.  Consequently, it is informative to consider a modified view of the connectivity mass based on pairwise mutual information outage.  This can be easily achieved by redefining $H_{ij}$ to be
\es{
    H_{ij}(r_{ij}) = P\left(\log_2(1+ g(r_{ij}) X_{ij} / \beta) \geq \zeta_m R \right)
\label{H2}}
where $R$ is a target rate threshold and $\zeta_m = 1$ if STBC is employed and $m \leq 2$ or if MIMO-MRC is considered, and $\zeta_m =2$ otherwise.  
Rearranging this it is clear that all previous calculations and results follow accordingly, but with $\wp$ replaced by $(2^{\zeta_m R} - 1)$ to explicitly account for the difference in rate characterised by $\zeta_m$.  Thus, we deduce from~\eqref{eq:Mdc} and~\eqref{eq:Mb} that for the rate based connectivity metric of \eqref{H2} we have that $M_{\text{dc}} < M_{\text{b}}$ since $(2^{2R}-1) > 2(2^R-1)$.

\textit{3) The case of $m/n\to y$:}
Now, let $m$ and $n$ scale such that their ratio approaches $y > 0$, then the relation $M_{\text{dc}} < M_{\text{b}}$ is maintained when considering mutual information outage \eqref{H2}. 
Neglecting the rate differences and focusing only on SNR outage (a proxy for reliability in delay-tolerant networks) as originally given in~\eqref{H1}, we see that when $y = (\sqrt 2 -1)^2 \approx 0.172$ the leading orders of the two schemes are equal. Denoting this critical value as $y_c$ we arrive at the conclusion that for $y < y_c$, diversity coding is favoured over beamforming to leading order, with the opposite being true for $y > y_c$. 
Again, we must be careful for finite $n$ since the correction terms differ. 
To draw further conclusions, we have plotted the connectivity mass against $n$ with $y \approx 0.172$ in the right panel of Fig.~\ref{fig1}.  
We see from Fig.~\ref{fig1} that diversity coding is preferred over beamforming for small~$n$ but not overly so.

We have thus shown that for the chosen metric of outage probability \eqref{H1} (reliability being the main concern relevant to delay tolerant networks – a paradigm of the IoT) beamforming results in superior network connectivity to diversity coding for $m=2$, and asymptotically when $y>y_c$. 
For the metric of pairwise mutual information \eqref{H2}, beamforming results in superior network connectivity to diversity coding for $m>2$.

\section{Design Implications}
The scaling laws developed herein can be exploited to mitigate the deleterious effects that boundaries present in confined networks using multiple antennas. 
Let our reference point be given by the connectivity mass $M_i$ corresponding to a homogeneous network connected by single-input single-output pairwise links, which can be computed to be $\Omega\Gamma(1+\mathcal C)/((\wp\beta)^{\mathcal C}d)$ using~\eqref{Mdc}.  In a bounded network, we can focus on a particular feature of solid angle $\omega$ and use, for example,~\eqref{eq:Mb} to obtain the \clred{antenna scaling} law that will ensure the effect that this feature has on local connectivity is mitigated.  Specifically, we see that
$wn \approx \big(\tfrac \Omega \omega \Gamma(1+\mathcal C)\big)^{1/\mathcal C}$, where $w = \zeta_m$ for diversity coding and $w = (1+\sqrt y)^2$ for beamforming.  Noting that these calculations pertain to nodes situated near boundaries, we can infer the possibility of designing sophisticated optimisation methods in practice.
Moreover, through the efficient utilisation of the derived scaling laws for the connectivity mass $M_i$, one can predict and indeed optimise both local (e.g. mean degree) and global (e.g. full connectivity) network performance.

\section{Conclusions}

In this letter we studied how scaling the per-node transmit power and the number of transmit/receive antennas $m/n$ affect both local and global network connectivity properties. 
By defining the \emph{connectivity exponent} $\mathcal C$ to be the ratio of the dimension of the network domain $d$ to the path loss exponent $\eta$, we showed that the connectivity mass $M_i$ controlling both local and global network properties scales like $\mathcal{O}(z^\mathcal C)$ in several parameters of interest $z$ (see for example equations \eqref{eq:MPT}, \eqref{Mdc}, \eqref{eq:Mb}).
Significantly, we analysed two MIMO signalling mechanisms (STBC and MRC) and showed that while both schemes scale like $\mathcal{O}( n^\mathcal{C})$, optimality depends on the relevant pre-factor and second-order correction terms.
We have also shown how antenna diversity scaling laws can be used to mitigate boundary effects in finite networks.


We have assumed negligible inter-node interference which for large scale ad hoc networks in low path loss environments becomes increasingly difficult to achieve.
Hence, interference effects must be included in future scaling law models and analysis, something which can be facilitated through the use of stochastic geometry tools \cite{Haenggi2009}.
Interestingly, very little work on bounded interference limited ad hoc networks has been published to date \cite{location}; a natural extension of the present work on MIMO transmission schemes and antenna diversity gain.

\section*{Acknowledgment}
The authors would like to thank the FP7 DIWINE project (Grant Agreement CNET-ICT-318177), and the directors of the Toshiba Telecommunications Research Laboratory.

\bibliographystyle{IEEEtran}
\bibliography{IEEEabrv,mybib,connectivity}

\end{document}